\pdfoutput=1
\documentclass[a4paper,fleqn]{cas-dc}

\usepackage[numbers,longnamesfirst]{natbib}

\def\tsc#1{\csdef{#1}{\textsc{\lowercase{#1}}\xspace}}
\tsc{WGM}
\tsc{QE}
\tsc{EP}
\tsc{PMS}
\tsc{BEC}
\tsc{DE}

\begin{document}
\let\WriteBookmarks\relax
\let\printorcid\relax 
\def\floatpagepagefraction{1}
\def\textpagefraction{.001}
\makeatletter\def\Hy@Warning#1{}\makeatother
\newcommand{\qh}{\textcolor{red}}

\shorttitle{Boosting Commit Classification with Contrastive Learning}

\shortauthors{Jiajun Tong et~al.}

\title [mode = title]{Boosting Commit Classification with Contrastive Learning}                      



%
\author[]{Jiajun Tong}[]





\affiliation[]{organization={China University of Mining and Technology, School of Computer Science and Technology},
    addressline={University Road}, 
    city={Xuzhou},
    postcode={221116}, 
    state={Jiangsu},
    country={China}}

\author[]{Zhixiao Wang}[]
\cormark[1]
\ead{zhxwang@cumt.edu.cn}
\author[]{Xiaobin Rui}[]



\cortext[cor1]{Corresponding author}


\begin{abstract}
Commit Classification (CC) is an important task in software maintenance, which helps software developers classify code changes into different types according to their nature and purpose. 
It allows developers to understand better how their development efforts are progressing, identify areas where they need improvement, and make informed decisions about when and how to release new software versions. However, existing models need lots of manually labeled data for fine-tuning processes, and ignore sentence-level semantic information, which is often essential for discovering the difference between diverse commits. Therefore, it is still challenging to solve CC in fewshot scenario.
  To solve the above problems, we propose a contrastive learning-based commit classification framework. Firstly, we generate $K$ sentences and pseudo-labels according to the labels of the dataset, which aims to enhance the dataset. Secondly, we randomly group the augmented data $N$ times to compare their similarity with the positive $T_p^{|C|}$ and negative $T_n^{|C|}$ samples. We utilize individual pretrained sentence transformers (ST)s to efficiently obtain the sentence-level embeddings from different features respectively. Finally, we adopt the cosine similarity function to limit the distribution of vectors, similar vectors are more adjacent. The light fine-tuned model is then applied to the label prediction of incoming commits. 
  Extensive experiments on two open available datasets demonstrate that our framework can solve the CC problem simply but effectively in fewshot scenarios, while achieving state-of-the-art(SOTA) performance and improving the adaptability of the model without requiring a large number of training samples for fine-tuning. The code, data, and trained models are available at \url{https://github.com/AppleMax1992/CommitFit}. 
\end{abstract}




\maketitle

\section{Introduction}
\label{sec:intro} 
During software development, developers use commits to track the change of codes within version control tools (such as GIT). 
Each time a developer makes changes to the codebase and commits those changes to the version control repository, they provide a description or message that explains what the changes are about.
Commit classification(CC) \citep{herivcko2023commit} is the process of categorizing individual code commits or changes made to a software project's version control system based on their purpose, intent, or content. 
The goal of commit classification is to group similar types of changes together, making it easier to understand their development progress, identify areas that require improvement, and make informed decisions regarding software version releases.
The challenge of the CC task is that the information in the commit message is usually not standardized, and a large number of commits are generated every day. Manual classification requires strong professional knowledge and consumes a lot of effort.

Previous works \citep{mockus2000identifying, zhou2017automated, levin2017boosting} investigate traditional machine learning methods to model commit message and their labels. Some of them divided commits into three categories, in which ``Corrective" for fixing faults, ``Perfective" for optimization of program process, and ``Adaptive" for applying new features. Some utilized static machine learning methods(e.g., Support Vector Machines (SVM) and XGBoost) for commit classification. This framework categorizes commits as either ``POSITIVE" for secure commits or ``NEGATIVE" for insecure commits.
\begin{figure}[ht]
  \centering
  \includegraphics[width=\linewidth]{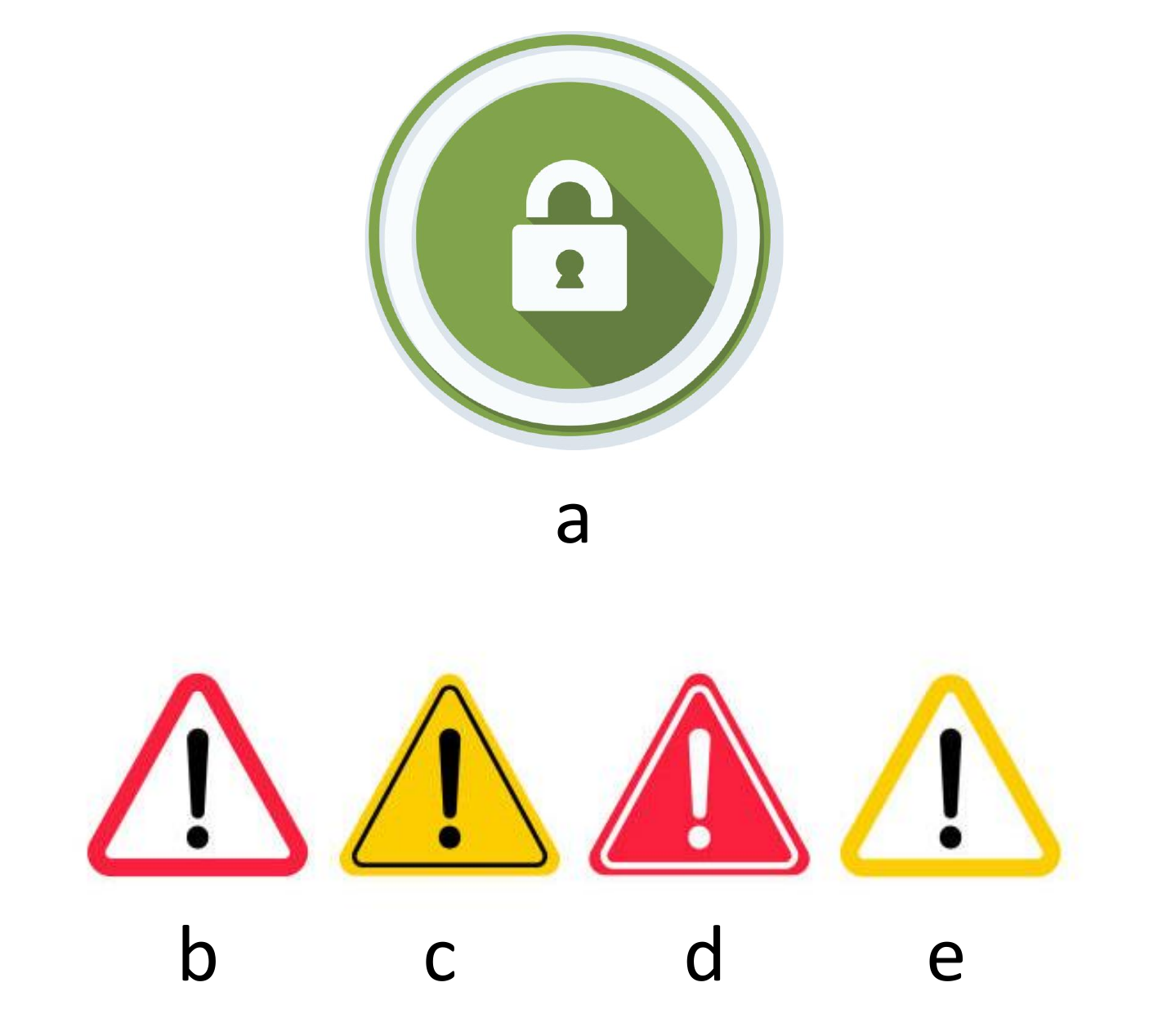}
  \caption{Motivation for incorporating contrastive learning on commit classification task. 
  As illustrated in this figure, even though there are only a small set of instances, and we don't know their exact label, we can early judge that \{b, c, d, e\} are similar by contrasting, while a is different from others. This motivates us to explore incorporating Contrastive learning on commit classification tasks.
  }
  \label{motivation for incorporating contrastive learning on commit classification task}
\end{figure}
Although these methods explore automated solutions for the CC problem, traditional machine learning models are not flexible enough and can only deal with fixed features.
Therefore, some works \citep{honel2019importance, wu2022enhancing} utilize the neural network to explore the features in the commit information adaptively. 
With the development of pre-trained models, researchers start to use pre-trained models to migrate the dependence on annotated commits.
\citep{ghadhab2021augmenting, sarwar2020multi} leverage prior knowledge encoded in pretrained models to enhance performance.
These methods effectively improve the accuracy of CC tasks, but those fully supervised approaches rely on large amounts of labeled data to train the model.
Recently, Lee et al. \citep{lee2021co} apply  co-training to semi-supervised learn the features from multiple views of the commits.
\begin{table*}[ht]
  \centering
  \caption{An example of the impact of sentence-level(SL) semantic information on commit prediction}
  \begin{tabular}{llll}
      Commit Message & True Label  & prediction & $prediction_{SL}$ \\ 
      \hline
      Fix null pointer exception in the login module & Corrective   & Corrective  & Corrective\\ 
      Refactor the database access layer for improved performance & Perfective  & Perfective & Perfective  \\ 
      Update user interface to support new screen resolutions & Perfective   & Adaptive & Perfective \\ 
      \hline
  \end{tabular}
  \label{sentence level feature}
\end{table*}

However, existing methods still require amounts of labeled data, making them not easily adaptable to fewshot scenarios. 
For instance, in actual industrial scenarios, the circle of maintenance is usually short, and some labels only have few labeled data. Therefore, it is difficult to have enough time to collect enough labels to start classification.
Moreover, existing methods ignore sentence-level information, which is often crucial for distinguishing the categories of commits. 
For example, if we have three commit samples for a software project as shown in Tab.\ref{sentence level feature}, without incorporating sentence-level embeddings, we might encounter a misclassification, the model might incorrectly classify the third commit as ``Adaptive'' instead of ``Perfective''. This error could occur because the words ``update" and ``interface" are often associated with ``Adaptive'' changes, suggesting modifications to accommodate external factors such as new devices or operating systems. In this case, the change is purely improving the existing user interface, which falls under the ``Perfective'' category.

To tackle above problems, we propose a framework for commit classification based on contrastive learning, which aims to learn useful representations by contrasting positive and negative pairs of data as shown in Fig.\ref{motivation for incorporating contrastive learning on commit classification task}. It leverages the inherent structure and relationships present in the data to train a model without relying on explicit labels. 
Firstly, we generate $K$ sentences and pseudo-labels based on the commit tags, which are used to augment the dataset. Secondly, we randomly group the augmented data $N$ times and compare their distances with the positive $T_p^{|C|}$ and negative $T_n^{|C|}$ samples. Next, we employ individual pretrained STs to efficiently extract sentence-level embeddings from different features, which is essential to distinguish sentences' similarity. Finally, we utilize the cosine similarity to constrain the vector distribution, ensuring that similar vectors are closer together. The fine-tuned model is then employed for predicting labels of incoming commits. Last but not least, for the first time, we conduct extensive experiments on two public datasets with different classification criteria to evaluate the effectiveness of our proposed method and its performance in fewshot scenarios. The experimental results show that our framework not only achieves the SOTA performance, but is also competitive in fewshot scenarios.


\begin{figure*}[ht]
  \centering
  \includegraphics[width=\linewidth]{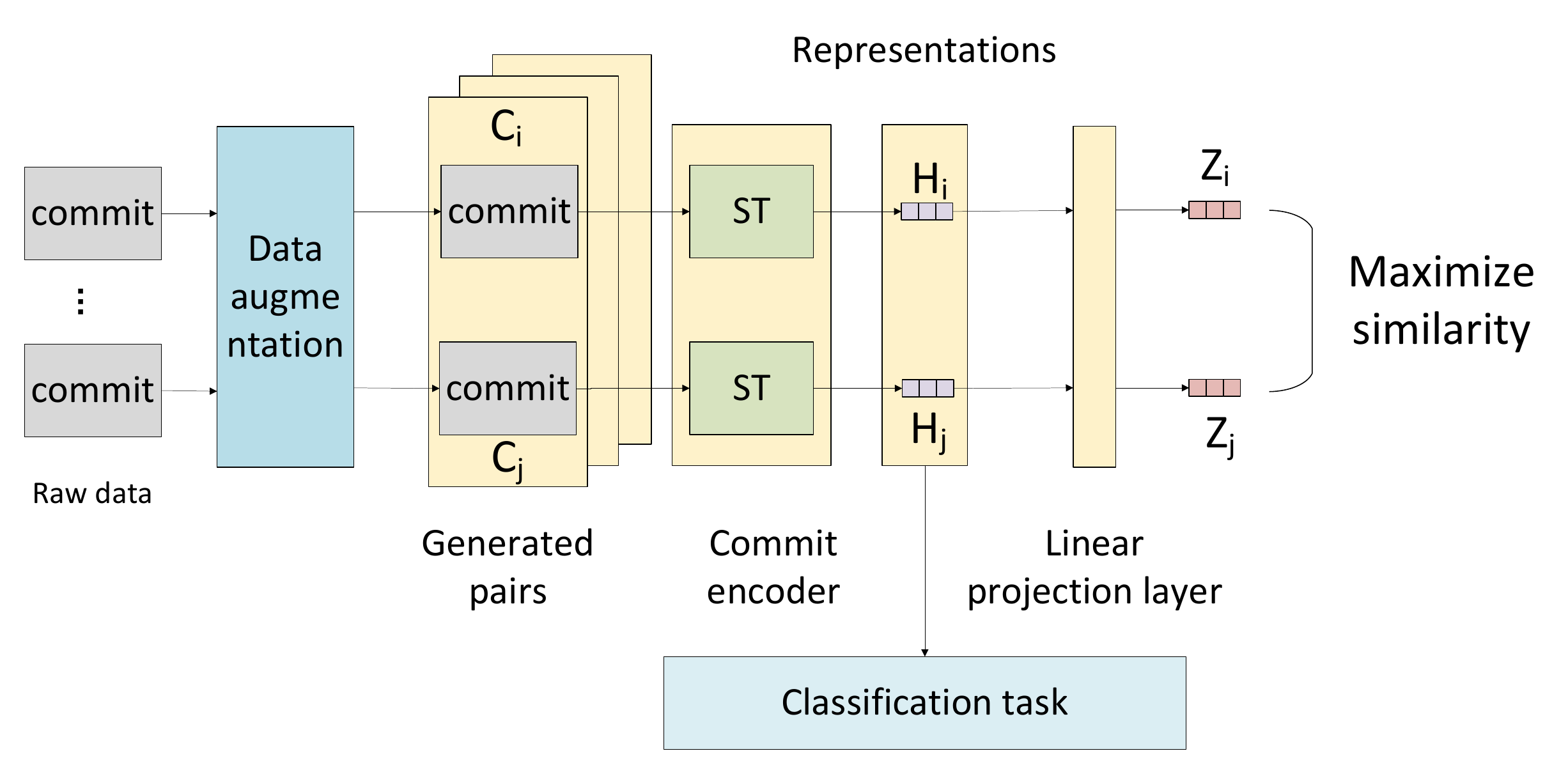}
  \caption{Overall view for \textbf{BooCC}. 
  a) the original data is used to generate an enhanced dataset through a template, and then $N$ groups are randomly divided into each commit. b) We use ST to embed multiple features in the commit, stitch the embedded vectors together, and then map the vectors to the same space through a linear transformation layer. c) We adopt the cosine similarity function to limit the distribution of vectors, similar vectors are more adjacent. d) Finally, the fine-tuned model is applied to the label prediction of incoming commits.
  }
  \label{Overall view for BooCC}
\end{figure*}

The main contributions of this work can be summarized as follows:
\begin{itemize}
  \item [1)] To address the challenge of limited labeled training data, we propose a contrastive learning based framework for commit classification. we generate $K$ sentences and pseudo-labels according to the labels of the dataset, which aims to enhance the dataset, and are randomly grouped $N$ times to compare the distance with their positive $T_p^{|C|}$ and negative $T_n^{|C|}$ samples.
  
  \item [2)] To efficiently learn sentence-level embeddings, we employ sentence-transformers (ST)s to obtain the vector representations from augmented datasets, and then self-supervised pull similar vector representations nearly by comparing the similarity of embedding vectors.
  \item [3)] We analyze and conduct extensive experiments on two public datasets to evaluate the effectiveness of our proposed model. The experimental results show that our framework demonstrates the SOTA performance on two CC dataset and have strong adaptability, while does not rely on complex fine-tuning procedures in a fewshot scenario. We made the data used and the proposed model public in \url{https://github.com/AppleMax1992/CommitFit}. 
\end{itemize}

\section{Related work}

\noindent
\textbf{Contrastive Learning}
\noindent
Contrastive Learning \citep{chen2020simple, jaiswal2020survey} has primarily been applied in the field of commit processing. It is utilized to learn common features from an unlabeled dataset by teaching the model to differentiate between similar and dissimilar data points. CL is widely used in the computer vision fields. 
Jung et al. \citep{jung2022exploring} introduce a method that combines semantic relation consistency (SRC) regularization and decoupled contrastive learning (DCL). This approach leverages diverse semantics by emphasizing the heterogeneous relationships between image patches within a single image. 

Yang et al. \citep{yang2022unified} present a novel learning approach known as Unified Contrastive Learning (UniCL). This method introduces a single learning objective that effectively promotes synergy between two types of data. The aim is to facilitate seamless integration of the two data types, leading to improved performance.
Recently, there have been a few attempts to extend contrastive learning to text classification tasks. For example, a contrastive learning system called ContrastNet is proposed by Chen et al. \citep{chen2022contrastnet} to address the issues of discriminative representation and overfitting in fewshot text categorization. For text classification applications, Pan et al.\citep{pan2022improved} suggest a straightforward and all-purpose technique to regularize the fine-tuning of Transformer-based encoders. However, to the best of our knowledge, no existing work has explored the application of contrastive learning specifically for commit classification. 

\noindent
\textbf{Pretrained Language Model}
The research on Pretrained Language Model(PLM) is a very popular trend now, both in the fields of CV \citep{perez2021true,kobyzev2023we} and NLP \citep{li2022pretrained,rethmeier2023primer}, which are trained on massive amounts of diverse text data, and aims to enables models to learn general representations on specific ML tasks with relatively limited labeled data. 
For example, to effectively inject knowledge adapters into the fundamental PLMs for fine-tuning the extractive summarization task, Xie et al.\citep{xie2022pre} investigate generative and discriminative training techniques to fuse domain knowledge (i.e., PICO elements).
BLIP-2 \citep{li2023blip} bootstraps vision-language pre-training using frozen big language models and commercially available pre-trained image encoders. 
Recently, the PLM tends to be larger and larger, (e.g., GPT \citep{lund2023chatting} and T5 \citep{sanh2021multitask}). These models are working well with multi-task and generative frameworks, but they are too large to fine-tune and naturally not suitable for classification tasks. Although scholars have tried to apply PLM for CC tasks \citep{ghadhab2021augmenting, lee2021co}, these models did not consider sentence-level embedding information, and did not pay attention to the problem of multilingualism in CC tasks, so we leverage sentence transformer as our pre-training model in this paper.

\noindent
\textbf{Commit Classification} 
Commit classification plays a crucial role in software maintenance as it helps developers effectively manage code and mitigate risks. Over the years, there has been growing research dedicated to this field. Mockus et al. \citep{mockus2000identifying} proposed significant definitions for commit classification, laying the foundation for subsequent studies. Several models \citep{zhou2017automated} have utilized static machine learning methods(e.g., Support Vector Machines (SVM) and XGBoost) for commit classification. For example, to automatically identify commits that are security-relevant, Sabetta et al. \citep{sabetta2018practical} propose an approach based on machine learning and analyze source code repositories. Mariano et al. \citep{mariano2019feature}  utilize XGBoost as the classifier with three additional features.
\citep{honel2019importance, wu2022enhancing} obtain the features in the commit information by the neural network adaptively. With the advancement of pretrained models, recent works \citep{lee2021co, ghadhab2021augmenting} leverage prior knowledge encoded in pretrained models to enhance performance. However, these methods often require lots of training samples and complex fine-tuning procedures, and most of them rely on supervised learning. The main limitation of supervised learning is the need for annotated data, which may not be readily available for many tasks. To overcome this challenge, we propose leveraging contrastive training as a means to address this issue within our framework. By incorporating contrastive learning, we aim to enhance the model's ability to generalize and perform well even with limited labeled training data.

\section{METHODOLOGY}

\subsection{Problem definitions}
The problem we address in this study is to predict the label $y$ of a given commit message $x$. Specifically, given a few labeled commits $(x_i, y_i)$, we adopt the commonly used $N$-way $K$-shot strategy to train the model, where $N$ represents the number of commit classes and $K$ represents the quantity of annotated commits for each class. Each task consists of a support set $S$ that contains $N \times K$ support instances and a query set $Q$. We train a classifier using the support set $S$ and evaluate its performance on the query set $Q$. In this setting, a higher value of $N$ and a lower value of $K$ indicate a more challenging task, as the classifier needs to generalize well with limited annotated data.

\subsection{Overview}
In this section, we present our proposed method \textbf{Boo}sting \textbf{C}ommit Classification with \textbf{C}ontrastive Learning(\textbf{BooCC}). We first give an overall view for BooCC as shown in Fig.\ref{Overall view for BooCC}. 
Then we explain the component of Data Augmentation. Further, we introduce the embedding model. Finally, we outline the training and inference process of our proposed method.

For each commit in the dataset, the template is first used to generate N sentences, and the dataset is augmented with known classification labels. We then want the model to know whether this arbitrary pair of commits in the augmented dataset is ``similar'' in that they are essentially different versions of the same classification. We can feed these two commits into our encoder model (a pretrained sentence transformer), creating a vector representation for each commit. For commits with multiple features, we utilize separate STs to embed different features separately, then concatenate them as the final embedding representation. Then we map the features of different dimensions to the same latent space, in which similar data points are expected to be close to each other, and the distance or proximity between points can indicate similarity or dissimilarity over training time, through a linear transformation layer.  For example, the commits labeled ``SECURE" should have similar representations, and the representation of ``INSECURE" should be different from the representation. Our goal is to train the model to distinguish commits between different types even without knowing what the true labels of the commits are. The similarity score is calculated by cosine similarity. Finally, we get a fine-tuned model with a projection head. In the inference stage, we can directly obtain the real label of the commit through the fine-tuned model through the embedded information of the commit. Overall, this contrastive learning approach can be split into three main components: Data Augmentation, Commit Representation, and Loss Function.

\begin{figure}[ht]
  \centering
  \includegraphics[width=\linewidth]{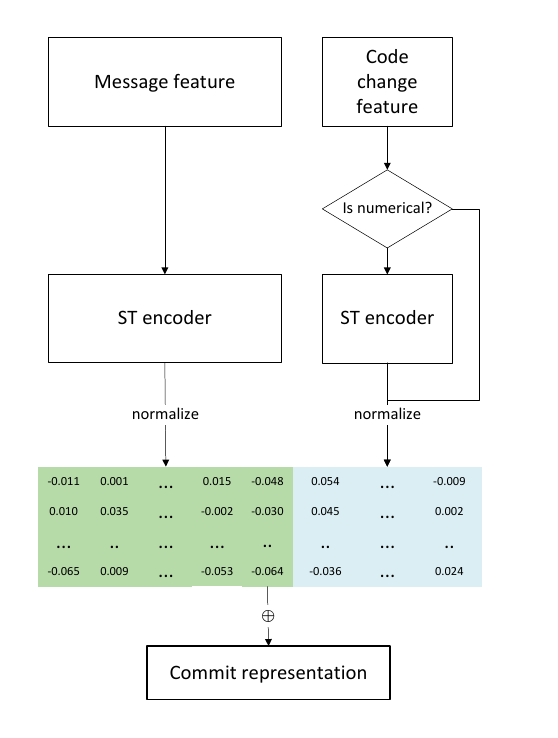}
  \caption{Commit Representation. 
  To efficiently obtain sentence-level embedding information, we here utilize ST as a commit encoder. Usually, the message features in the commit are the same, which are the natural language comments by developers from different countries. However, code change information is different in the two datasets. Among them, the three-category dataset already has standardized numeric data with 0 and 1, which can be directly concatenated, while the code change information in the two-category is the original programming language, so we use an additional ST for its embedding.
  }
  \label{Commit Representation}
\end{figure}
\subsection{Data Augmentation}
To better handle the limited amount of labeled training data in fewshot scenarios, we adopt a contrastive training approach that is often used for image similarity. Given a small labeled training set $D = \{(x_i, y_i)\}$, where $x_i$ represents sentences and $y_i$ represents their corresponding class labels, we aim to generate additional training samples for better generalization. For each class label $c \in C$, we first create $N$ samples by a template ``This sentence is '' $c$ to extract the most superficial features from each class. We further build a set of positive triplets $T_p^c = \{(x_i, x_j, 1)\}$, where $x_i$ and $x_j$ are randomly selected sentence pairs from the same class $c$. In other words, $(y_i = y_j = c)$. Similarly, we build a set of negative triplets $T_n^c = \{(x_i, x_j, 0)\}$, where $x_i$ represents sentences from class $c$, and $x_j$ represents randomly chosen sentences from different classes such that $(y_i = c, y_j \neq c)$. By combining the positive and negative triplets across all class labels, we form the contrastive dataset $T$. Mathematically, $T$ is defined as:
\begin{equation}
T = \{(T_p^0, T_n^0), (T_p^1, T_n^1), ..., (T_p^{|C|}, T_n^{|C|})\}
\end{equation}
where $|C|$ represents the number of class labels, and $|T| = 2R|C|$ denotes the total number of pairs in $T$. Here, $R$ is a hyperparameter that controls the number of positive and negative triplets generated for each class. In our evaluations, we set $R = 20$ unless specified otherwise.

\subsection{Commit Representation}
To encode an input sentence into a dense vector representation, we pass input sentences to a pre-trained model that converts them into fixed-length vectors that capture semantic and contextual information as shown in Fig. \ref{Commit Representation}. Specifically, we add the fine-tuned ST to encode each feature $h_i^{msg}$, $h_i^{cc}$ of the original labeled training data ${x_i}$, then we normalize the embedded features through the $L_p$ norm operation. For a embedding $h$ of size $k$ ($n_0$, $n_1$,...,$n_k$), each $n_dim$-elements vector 
$h$ along dimension $dim$ can be transformed as 
$ \hat{h} = \frac{h}{max({||h||}_p,\epsilon)}$ where $p$ is the exponent value in the norm formulation, $dim$ is the dimension to reduce and $\epsilon$ is the small value to avoid division by zero. Then we concatenate the normalized features to obtain the embedded representation of commit $H_i = \hat{h}_i^{msg} \bigoplus \hat{h}_i^{cc}$, where $\bigoplus$ represents the concatenation operation between $\hat{h}_i^{msg}$ and  $\hat{h}_i^{cc}$.
Furthermore, we apply a logistic regression as an additional transformation layer to generate submission projection heads, which are calculated as:
\begin{equation}
  \begin{split}
  z_{v}& =z_{s}\hat{H_j} \\ 
       &=\frac{H_i\cdot H_j}{\left \| H_j \right \|} \frac{H_j}{\left \| H_j \right \|} \\
       &= \frac{H_i \cdot H_j}{H_j\cdot H_j}H_j 
  \end{split}
\end{equation}
where $z_{s}$ yields:
\begin{equation}
  \begin{split}
  z_{s} &= \left \| H_i \right \|cos\Theta \\ 
  & = \left \| H_i \right \|\frac{H_i \cdot H_j}{\left \| H_i \right \|\left \| H_j \right \|} \\
  & = \frac{H_i \cdot H_j}{\left \| H_j \right \|} 
\end{split}
\end{equation}
where $H_i$ is the source vector for projection, and $H_j$ is the target for projection.  While $\hat{H_j}$ is the unit vector in the direction of $H_j$ and $z$ is the project of $H_i$ onto $H_j$. And $z_s$ is the scalar projection of $H_i$ on $H_j$. $z_v$ is the vector projection of $H_i$ on $H_j$. $||||$ is the length of a vector and $\cdot$ representation the dot product operation. $\Theta$ is the angle between $a$ and $H_j$ vectors.


\subsection{Loss Function}
To minimize the loss during training, we adopt the normalized temperature-scaled cross-entropy loss (NT-Xent loss) as the loss function. First, we compute the cosine similarity between two commits by projecting representations $z_i$ and $z_j$.
We calculated the pairwise similarity between two commits as: 
\begin{equation}
  sim_{i,j} = \frac{{z_{i}^{T}z_{j}} }{(\tau||{z_{i}}|| ||{z_{j}}||)}
\end{equation}
Where $||z_||$ is the norm of the vector. And $\tau$ stands for the adjustable temperature parameter, which scales the inputs to the range [-1, 1] of cosine similarity. The temperature parameter $\tau$ controls the spread of the probability distribution. A higher value of $\tau$ makes the probabilities more uniform, while a lower value concentrates the probabilities on the most similar commit pairs. 
We then adopt the softmax function to calculate the likelihood that two commits (represented as $i$ and $j$) are comparable, and then we apply the Noise Contrastive Estimation (NCE) Loss to train the model. The NCE Loss helps us to distinguish positive pairs (comparable commits) from negative pairs (non-comparable commits).

The formula for the NCE Loss is given as follows:

\begin{equation}
  l(i, j) = -log\frac{exp(sim_{i, j})}{ \sum_{k=1}^{2N} \mathbb{I}_{[k!= i]} exp(sim_{i, k})}
\end{equation}

In this formula, the term $sim_{i, j}$ represents the similarity score between commits $i$ and $j$, which is computed using some similarity metric (e.g., cosine similarity) based on their representations. The denominator $\sum_{k=1}^{2N} \mathbb{I}{[k \neq i]} \exp(sim{i, k})$ is the sum of the similarity scores between commit $i$ and all other commits in the batch, excluding itself ($k \neq i$). This term acts as a normalization factor to scale the probabilities and make them sum up to 1. While the indicator function $\mathbb{I}_{[k \neq i]}$ is used to evaluate whether the commits $i$ and $k$ are different. It takes the value 1 when $k$ is not equal to $i$ (i.e., $k \neq i$), and 0 otherwise. This ensures that we exclude the similarity score of the commit with itself in the denominator.
In the final step, we compute the loss over all pairs in the batch of size $N$ to obtain an average score. The loss function is defined as follows:
\begin{equation}
  L = \frac{1}{2N} \sum_{k=1}^{N} [l(2k-1, 2k) + l(2k, 2k-1)]
\end{equation}
where $l(\cdot)$ represents the loss function used to compare the representations of pairs of samples. The encoder and projection head representations improve over time based on this loss, resulting in representations that bring related commits closer together in the space.
During the inference stage, instead of using the projection head $g(\cdot)$, we utilize the encoder $f(\cdot)$ and its corresponding representation $h$ for subsequent tasks. This ensures that the refined and meaningful representations obtained during the training process are employed for further tasks or applications.
\begin{table*}[!ht]
  \caption{Samples of Dataset I}
  \resizebox{\textwidth}{!}{\begin{tabular}{llll}
      Samples of Dataset I & ~ & ~ & ~ \\ 
      \hline
      Commit\_ID & Project & Comment & 3\_labels \\ 
      \hline
      0531b8b & ReactiveX-RxJava & Change hasException to hasThrowable-- & p \\ 
      013fd99 & hbase & \makecell[l]{Alter table add cf doesn't do- compression test (Virag Kothari)--} & c \\
      48d33ec & elasticsearch ~ & support yaml detection on char sequence-- & a \\ \hline
  \end{tabular}
  }
  \label{tab: Samples of Dataset I}
\end{table*}
\begin{table*}[!ht]
  \caption{Samples of Dataset II}
  \resizebox{\textwidth}{!}{ \begin{tabular}{lllll}
      Samples of Dataset II & ~ & ~ & ~ & ~ \\ 
      \hline
      Github & Message & Diff & Label \\ 
      \hline
      https://github.com/gpac/gpac/commit/e115e3bbdb... & added ignore list when checking unused args  & diff --git a/include/gpac/filters.h b/include/... & 1 \\ 
      https://github.com/axiomatic-systems/Bento4/co... & allow tracks with different frame rates in the...  & diff --git a/Source/Python/utils/mp4-dash.py b... & 0 \\ 
      https://github.com/denkGroot/Spina/commit/4515... & Translations including fallbacks (\#430)$\backslash$n$\backslash$n* T...  & diff --git a/app/controllers/spina/admin/pages... & 1 \\ 
      \hline
  \end{tabular}
  }
  \label{tab: Samples of Dataset II}
\end{table*}
\section{Experiment}
In this section, we first introduce the dataset and the experiment setup. Then we compared several baselines to verify the effectiveness of our model. Further, we evaluate the performance of different pretrained models on fewshot scenarios. Finally, we discuss the time complexity of the model to illustrate the efficiency of the model. We provide the hyper-parameter values for \textbf{BooCC} as follows: We utilize the default parameters from Huggingface \footnote{\url{https://huggingface.co}} to ensure fairness. Specifically, we introduce AdamW as the optimizer function, with a learning rate of 1e-5 and the batch size is set to 64. And we limit the early stop function up to 10 epochs after no improvements on Acc. All models are trained on Gpushare Cloud a leading GPU Cloud service provider from China \footnote{\url{https://gpushare.com/}} instance 24GB Nvidia 3090 and Intel(R) Xeon(R) CPU E5-2683 v4 with 40GB memory. 
The detailed environment information is shown as Tab.\ref{tab: Env Info}.
\begin{table}[!ht]
  \caption{Env Info}
  \begin{tabular}{|l|l|}
  \hline
      transformers\_version & 4.30.2 \\ \hline
      framework & PyTorch \\ \hline
      use\_torchscript & False \\ \hline
      framework\_version & 2.0.0+cu118 \\ \hline
      python\_version & 3.8.10 \\ \hline
      system & Linux \\ \hline
      cpu & x86\_64 \\ \hline
      architecture & 64bit \\ \hline
      use\_multiprocessing & True \\ \hline
      only\_pretrain\_model & False \\ \hline
      cpu\_ram\_mb & 128801 \\ \hline
      use\_gpu & True \\ \hline
      num\_gpus & 1 \\ \hline
      gpu & NVIDIA GeForce RTX 3090 \\ \hline
      gpu\_ram\_mb & 24576 \\ \hline
      gpu\_power\_watts & 350.0 \\ \hline
      gpu\_performance\_state & 0 \\ \hline
      use\_tpu & False \\ \hline
  \end{tabular}
  \label{tab: Env Info}
\end{table}


  \subsection{Experimental Setup} 

  \noindent
  \textbf{Data Acquisition and Processing}
  \noindent
  We evaluate two publicly available datasets to demonstrate the effectiveness of our proposed model.  Since ``comment'' is the common attribute on the two datasets, we want to demonstrate our model as a simple and general approach. Therefore, we choose comments as our only input. Ghadhab et al. \citep{ghadhab2021augmenting} combined three datasets \citep{mauczka2015dataset, mauczka2015dataset, alomar2019can} which collect commits from open-source projects that cover several domains (e.g., databases, programming languages, and integration frameworks), with 1,793 annotated commits and three categories for software maintenance activities identification. We denoted it as Dataset I\footnote{\url{https://zenodo.org/record/4266643\#.X6vERuLPxPY}} and present the data characteristics in Tab.\ref{tab: Data charcteristics of Dataset I} and Tab.\ref{tab: Samples of Dataset I}. 
  Lee et al. \citep{lee2021co} create a dataset from RA-Data \citep{reis2021ground}, which consists of 3,765 positive samples and roughly 6,300 negative samples from 910 repositories with two classes SECURE and INSECURE. We denoted it as Dataset II\footnote{\url{https://github. com/davidleejy/wnut21-cotrain}} and present the data characteristics in Tab.\ref{tab: Data characteristics of Dataset II} and Tab.\ref{tab: Samples of Dataset II}.

  \begin{table}[!ht]
    \centering
    \caption{Data charcteristics of Dataset I}
    \begin{tabular}{ll}
        Data Characteristics of Dataset I & ~ \\ \hline
        Category label & Number of Instances \\ \hline
        Corrective & 600 \\
        Adaptive & 590 \\ 
        Perfective & 603 \\ 
        \hline
        Total & 1793 \\ 
        \hline
    \end{tabular}
    \label{tab: Data charcteristics of Dataset I}
\end{table}

  \begin{table}[ht]
    \centering
    \caption{Data Charcteristics of Dataset II}
    \begin{tabular}{ll}
        Data Charcteristics of Dataset II & ~ \\ 
        \hline
        Category label & Number of Instances \\ 
        \hline
        Positive & 3765 \\ 
        Negative & 6347 \\ 
        \hline
        Total & 10112 \\ 
        \hline
    \end{tabular}
    \label{tab: Data characteristics of Dataset II}
  \end{table}
  We split the datasets with 70\% train, 15\%test, and 15\% validation. For fewshot scenarios, we follow the general N-way K-shot sampling strategy to extract training samples from the training dataset.
  
  \noindent
  \textbf{Baselines}
  In our evaluation, we compare our approaches  with two typical commit classification methods. Although there are many articles on committing classification, few of them are discussed under openly available datasets. To verify our method on this public dataset, we choose two methods to evaluate our methods' effectiveness as follows:
  \begin{itemize}
    \item $DNN@BERT+Fix\_cc$: Ghadhab et al. \citep{ghadhab2021augmenting} introduced a DNN model which concatenates the BERT-based word embeddings of commit messages and source code changes, and released a 3-way dataset.
    \item CR-ds: Lee et al. \citep{lee2021co} treated code changes and commit messages as two different views and used CodeBERT and RoBERTa to process them, respectively. They applied co-training to jointly train the two models. It is a 2-way task, which aims to identify security-related commits.
    \item  $BooC_{mge}$: We will only use the method of message information as our basic method $BooC_{mge}$. Since both data sets contain message information, according to the argument of Levin et al. \citep{levin2017boosting}, the message is the most instructive information for classification.
    \item $BooC_{mge+cc}$: Since both datasets contain cc information, we build $BooC_{mge +cc}$ to leverages message information and code change information together.
    \item $BooC_{mge}^{RoBERTa}$: It is the solution with the best classification effect, using a larger pre-training model, but to a certain extent lost the time of model training and mapping.
    \item $BooC_{mge+cc}^{RoBERTa}$: The best solution considering the speed and classification effect.
  \end{itemize}
  \noindent
  \textbf{Metrics} 
  \begin{figure*}[ht]
      \centering
      \includegraphics[width=\linewidth]{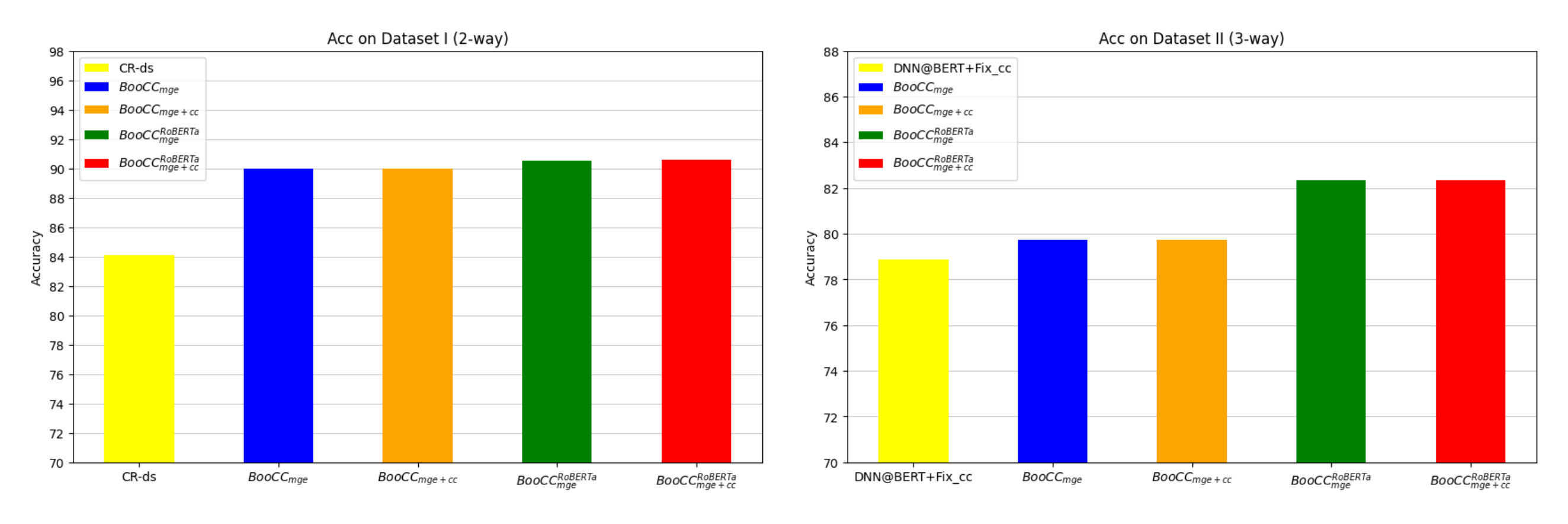}
      \caption{Acc performance on different datasets.
      }
      \label{Acc performance on different datasets}
  \end{figure*}
  We here leverage Precision, Recall, and F1-score as evaluation metrics to provide a comprehensive assessment of the model's performance in handling imbalanced and multiclass classification tasks, which are common characteristics of commit classification problems.
  Precision measures the proportion of true positive samples among all the samples that the model predicted as positive. It focuses on the accuracy of positive predictions, making it essential when we want to minimize false positives. In the context of commit classification, precision would help us understand how many of the commits predicted as a certain class are actually relevant and correct.
  \begin{equation}
    Precision = \frac{true\_positives}{true\_positives + false\_positives}
  \end{equation}
  Recall, also known as sensitivity or true positive rate, measures the proportion of true positive samples among all the actual positive samples in the dataset. It focuses on avoiding false negatives and capturing as many positive samples as possible. In commit classification, recall would indicate how effectively the model identifies all commits belonging to a specific class.
  \begin{equation}
    Recall = \frac{true\_positives}{true\_positives + false\_negatives}
  \end{equation}
  The F1 score is the harmonic mean of precision and recall. It combines both metrics and provides a balanced evaluation of the model's performance. The F1 score is particularly useful when there is an imbalance between the classes in the dataset. In commit classification, where some classes may have significantly fewer examples than others, the F1 score can give us a more reliable performance measure than using accuracy alone.
  \begin{equation}
    F1-score = 2 * \frac{precision * recall}{precision + recall}
  \end{equation}
  These metrics are widely used in the field of machine learning, especially in tasks with imbalanced datasets, such as commit classification. By considering both precision and recall, the F1 score allows us to assess the model's performance more comprehensively, taking into account false positives and false negatives and providing a more informative evaluation of the model's ability to classify commits correctly.
  Moreover, we measure the proportion of correctly classified samples (both true positives and true negatives) among all the samples in the dataset through Accuracy (Acc), which is another commonly used evaluation metric in commit classification and other machine learning tasks and commonly used to evaluate the performance of models \citep{li2022survey,li2020survey} in the fewshot scenario.
  The formula for Accuracy is as follows:
  \begin{equation}
      Accuracy = \frac{true\_positives + true\_negatives}{total\_samples}
  \end{equation}
  In the context of commit classification, accuracy provides a general overview of how well the model is performing in terms of correctly predicting the class labels across all classes. It is a straightforward and easy-to-understand metric that is often used as a primary performance indicator.
  \begin{figure*}[ht]
    \centering
    \includegraphics[width=\linewidth]{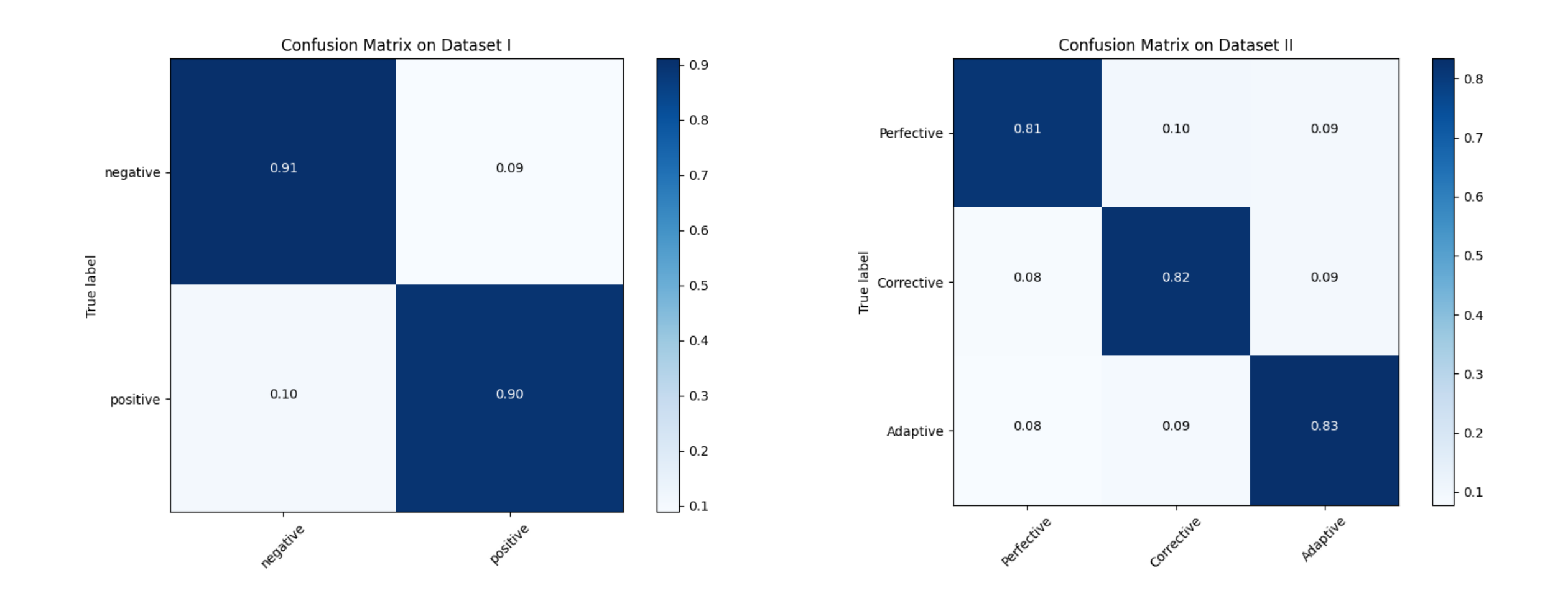}
    \caption{Confusion matrix on different datasets. 
    }
    \label{Confusion matrix}
  \end{figure*}
  \begin{table}[ht]
    \centering
    \caption{Performance comparison on Dataset I} 
    \begin{tabular}{llll}
        Approach & \multirow{2}{*}{P} & \multirow{2}{*}{R}& \multirow{2}{*}{F} \\ 
        Training set 7000+ \\ \hline
        CR-ds & 84.0 & 84.1 & 83.9 \\
        $BooCC_{mge}$ & 90.17 & 90.01 & 90.06 \\
        $BooCC_{mge+cc}$ & 90.07 & 90.01 & 90.04 \\
        $BooCC_{mge}^{RoBERTa}$ & 90.68 & 90.54 & 90.58 \\
        $BooCC_{mge+cc}^{RoBERTa}$ & \textbf{90.71} & \textbf{90.57} & \textbf{90.61} \\ \hline
        Training set 2*150 \\ 
        $\mathbf{BooCC_{mge+cc}^{RoBERTa}}$ & \textbf{84.76} & \textbf{84.88} & \textbf{84.75} \\ 
    \end{tabular}
    \label{tab: Performance comparison on Dataset I}
  \end{table}
  \begin{table}[ht]
    \centering
    \caption{Performance comparison on Dataset II} 
    \begin{tabular}{llll}
        Approach & \multirow{2}{*}{P} & \multirow{2}{*}{R}& \multirow{2}{*}{F} \\ 
        Training set 1200+ \\ \hline
        $DNN@BERT+Fix\_cc$ & 80.0 & 79.7 & 79.7 \\  
        $BooCC_{mge}$  & 79.76 & 79.74 &  79.74 \\
        $BooCC_{mge+cc}$ & 79.76 & 79.74 & 79.74 \\
        $BooCC_{mge}^{RoBERTa}$ & 82.35 & 82.34 & 82.34 \\
        $BooCC_{mge+cc}^{RoBERTa}$ & \textbf{82.35} & \textbf{82.34} & \textbf{82.34} \\ \hline
        Training set 3*100 \\ 
        $\mathbf{BooCC_{mge+cc}^{RoBERTa}}$& \textbf{81.14} & \textbf{80.86} & \textbf{80.87} \\ \hline
    \end{tabular}
    \label{tab: Performance comparison on Dataset II}
  \end{table}
\subsection{Performance Comparison}
\label{sec: Performance Comparison}

To verify the comprehensive performance of our method when dealing with CC problems, we first use the same as in the literature on the two data sets, that is, 70\% of the data as the training set. We conducted test experiments on all the remaining datasets. Since the previous methods on datasets with two labels and three labels are independent, we compared the two datasets with different baselines respectively. 

On Dataset I. We can see that from Tab.\ref{tab: Performance comparison on Dataset I}, compared to the CR-ds method using message and code change features, our method $BooCC_{mge}$ has reached SOTA with 90.17\% precision, 90.01\% recall and 90.06\% f1-score when only MGE is considered.
Further from Fig.\ref{Acc performance on different datasets}, we can see that our method $BooCC_{mge}$ has reached 90.01\% Acc, while the other three baselines have hardly changed. This shows that our comparative learning framework has achieved a good impact, while the code change feature has only a weak effect,
which is consistent with \cite{levin2017boosting} argument, the commit message feature is more important than the source code change feature.

And further illustrate the necessity of applying contrastive learning and sentence-level features to improve the accuracy of CC prediction.
However, we found that compared with $BooCC_{mge}$ based only on the message feature, the $BooCC_{mge+cc}$ method that added the code change feature did not achieve better results with 90.07\% precision, 90.01\% recall and 90.04\% f1-score, and even dropped slightly. It may be caused by the lack of programming language training corpus in paraphrase-mpnet-base-v2.
While $BooCC_{mge}^{RoBERTa}$ uses the Roberta corpus as the pre-training model's ST to achieve better results with 90.68\% precision, 90.54\% and 90.58\%. Since the message information contains comment messages written by programmers from different countries, and the pre-training corpus is richer. The pre-trained model is more likely to achieve better results.
Finally, $BooCC_{mge+cc}^{RoBERTa}$ adopt the ST using the all-roberta-large-v1 as the pre-training model and considering both the message feature and code change features have achieved the best performance with 90.71\% precision, 90.57\% recall, and 90.61\% f1-score,  
This further illustrates the code change feature has a certain impact on the effectiveness of the model. 
Moreover, our method only trained on the 100shot scenario has shown competitive performance compared to the previous SOTA method. 
This briefly proves the feasibility of our method in the fewshot scenario, which we will verify in detail in the next section.

On Dataset II, we compare with $DNN@BERT+Fix\_cc$, which are trained on over 1200 samples and exploit message features and code change features. From Tab. \ref{tab: Performance comparison on Dataset II}, it can be seen that when only message features are considered, our method $BooCC_{mge}$ almost achieves the same effect, 79.76\% precision, 79.74\% recall and 79.74\% f1-score, which shows that message features are the key to improve the performance.
$BooCC_{mge}^{RoBERTa}$ further achieved the best effect with 82.35\% precision, 82.34\% recall, and 82.34\% f1-score. However, we found that $BooCC_{mge+cc}$ and $BooCC_{mge+cc}^{RoBERTa}$ did not improve as expected after adding cc features. This may be because, on the second dataset, the code change feature is already a numeric matrix, and such features can no longer show obvious differences under the premise of a large number of datasets. We will further discuss the effect of code change on the fewshot datasets in the next section. Nevertheless, similar to the results on Dataset I, our model still achieves very competitive performance when only using 100 samples per class, with 81.14\% precision, 80.86\% recall and 80.87\% f1-score. 
Further, we can see from Fig.\ref{Acc performance on different datasets} that, similar to the performance on Dataset I, our methods of adding code change features did not show delightful results, while the model based on RoBERTa corpus pretrained ST has the best Acc 82.34\%.

Eventually, we visualize the confusion matrix to show the overall Acc performance of the model on each type of label. As can be seen from Fig.\ref{Confusion matrix}, our model reached 90\% and 91\% respectively on the positive and negative labels of Dataset I, and  81\%, 82\% and 83\% respectively on the Perfective, Corrective, and Adaptive labels of  Dataset II. We can conclude that our model can be used universally for the current CC tasks, and has achieved the effect of SOTA, our model has shown strong adaptability, and only In the case of a few samples of datasets, it still shows strong competitiveness.

\begin{table*}[ht]
  \begin{center}
  \caption{BooCC's performance on Fewshot scenerios} 
 
  \begin{tabular}{cccccccccccccc}
    \hline
    \multirow{2}{*}{Shots} & \multirow{2}{*}{Methods} & \multicolumn{4}{c}{Dataset I} & \multicolumn{4}{c}{Dataset II} \\
    ~ & ~ & P & R & F & Avg.(acc) & P & R & F & Avg.(acc)  \\ \hline
    \multirow{4}{*}{5} & $BooCC_{mge}$ & 67.69 & 68.44 & 67.71 & 68.44 & \textbf{51.44} & \textbf{49.83} & \textbf{49.73} & \textbf{49.83} \\ 
    ~& $BooCC_{mge+cc}$ & \textbf{67.87} & \textbf{68.89} & \textbf{67.73} & \textbf{68.89} & 51.44 & 49.83 & 49.73 & 49.83 \\ 
    ~& $BooCC_{mge}^{RoBERTa}$ & 63.60 &64.60 & 63.69 & 64.60 & 43.88 & 43.92 & 43.47 & 43.92 \\ 
    ~& $BooCC_{mge+cc}^{RoBERTa}$ & 63.79 & 65.27 & 63.83 & 65.27 & 43.88 & 43.92 & 43.47 & 43.92  \\ \hline
    \multirow{4}{*}{10} & $BooCC_{mge}$ & \textbf{72.49} & \textbf{73.05} & \textbf{72.23} & \textbf{73.05} & 67.13 & 67.01 & 66.63 & 67.01 & \\ 
    ~& $BooCC_{mge+cc}$ & 72.21 & 72.82 & 71.90 & 72.82 & \textbf{67.37} & \textbf{67.23} & \textbf{66.87} & \textbf{67.24} \\ 
    ~& $BooCC_{mge}^{RoBERTa}$ & 69.42 & 70.17 & 69.49 & 70.17 & 65.58 & 65.69 & 65.47 & 65.69 \\ 
    ~& $BooCC_{mge+cc}^{RoBERTa}$ & 69.47 & 70.31 & 69.39 & 70.31 & 65.58 &65.69 & 65.47 & 65.69 \\ \hline
    \multirow{4}{*}{15} & $BooCC_{mge}$ & 75.53 & 75.90 & 75.50 & 75.90 & 69.39 & 69.43 & 69.08 & 69.43 \\ 
    ~& $BooCC_{mge+cc}$ & \textbf{75.55} & \textbf{75.94} & \textbf{75.54} & \textbf{75.94} & \textbf{70.04} & \textbf{70.06} & \textbf{69.75} & \textbf{70.06} \\ 
    ~& $BooCC_{mge}^{RoBERTa}$ & 73.51 & 73.71 & 73.59 & 73.71 & 66.90 & 67.01 & 66.66 & 67.01 \\ 
    ~& $BooCC_{mge+cc}^{RoBERTa}$ & 73.66 & 73.86 & 73.74 & 73.86 & 66.90 & 67.01 & 66.66 & 67.01 \\ \hline
    \multirow{4}{*}{20} & $BooCC_{mge}$ & 76.59 & 76.46 & 76.52 & 76.46 & 70.40 & 70.49 & 70.36 & 70.49 \\ 
    ~& $BooCC_{mge+cc}$ & \textbf{76.81} & 76.76 & \textbf{76.79} & 76.76 & 70.51 & 70.62 & 70.47 & 70.62  \\ 
    ~& $BooCC_{mge}^{RoBERTa}$ & 76.59 & 76.63 & 76.61 & 76.63 & \textbf{71.38} & \textbf{71.31} & \textbf{71.28} & \textbf{71.31} \\ 
    ~& $BooCC_{mge+cc}^{RoBERTa}$ & 76.74 & \textbf{76.77} & 76.76 & \textbf{76.77} & 71.38 & 71.31 & 71.28 & 71.31 \\ \hline
    \multirow{4}{*}{50} & $BooCC_{mge}$ & 79.84 & 79.81 & 79.82 & 79.81 & 74.04 & 73.94 & 73.96 & 73.94 \\ 
    ~& $BooCC_{mge+cc}$ & \textbf{80.11} & \textbf{80.01} & \textbf{80.06} & \textbf{80.01} & 73.97 & 73.88 & 73.89 & 73.88 \\ 
    ~& $BooCC_{mge}^{RoBERTa}$ & 79.41 & 79.27 & 79.33 & 79.27 & \textbf{76.47} & \textbf{76.36} & \textbf{76.41} & \textbf{76.36} \\ 
    ~& $BooCC_{mge+cc}^{RoBERTa}$ & 79.43 & 79.29 & 79.35 & 79.29 & 76.47 & 76.36 & 76.41 & 76.36 \\ \hline
  \end{tabular}
  
  \label{tab: Performance on Fewshot scenerios}
\end{center}
\end{table*}

\begin{figure*}[ht]
  \centering
  \includegraphics[width=\linewidth]{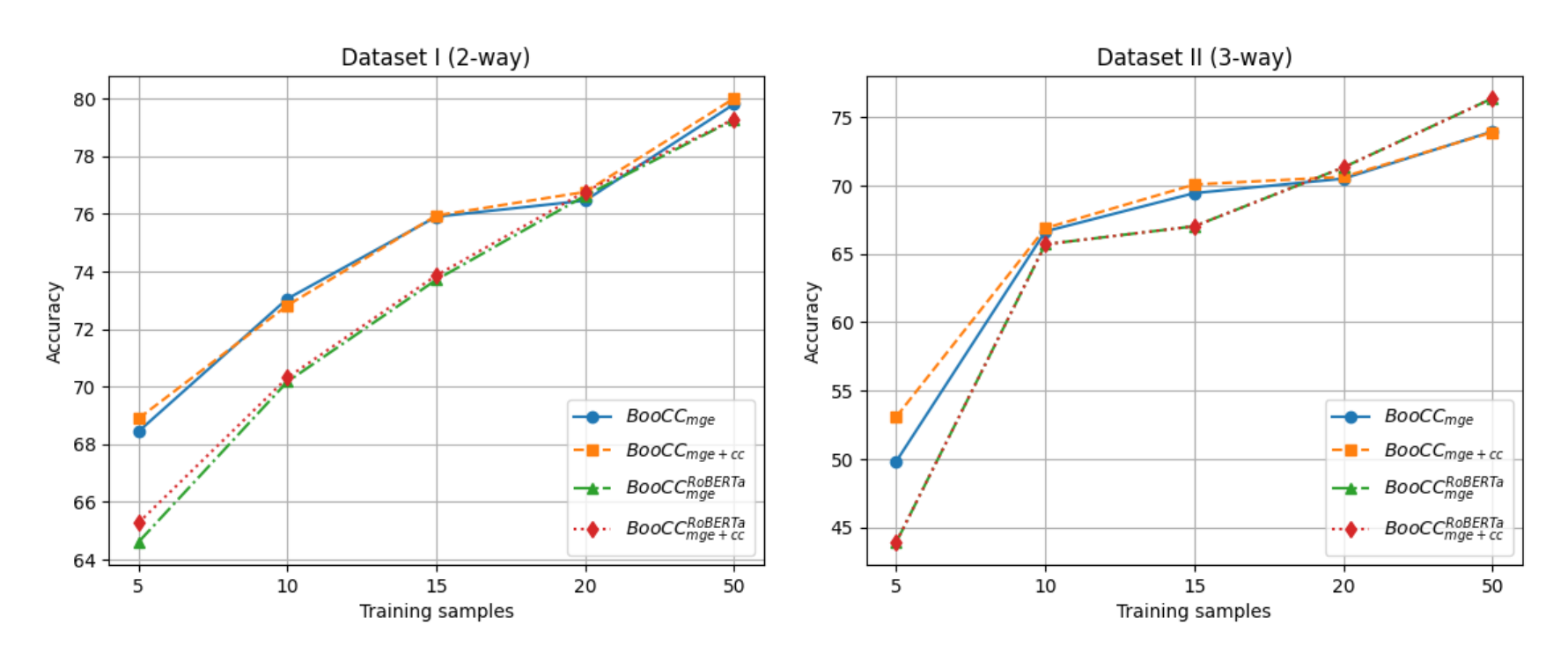}
  \caption{Fewshot performance on different datasets
  }
  \label{fig: Fewshot perfermance}
\end{figure*}
\subsection{The Effectiveness of BooCC on Fewshot}
\label{sec: BooCC on fewshot}
To verify the performance of our model in the fewshot scene, we trained 4 baselines with \{5, 10, 15, 20, and 50\} shots on the two datasets for comparison.
In order to be closer to the real industrial scene, we divide the dataset in a completely random way and keep them unbalanced. We utilize all the rest data except the training samples as the test set. And We have bolded the best test results in each case. From Tab. \ref{tab: Performance on Fewshot scenerios} we can see that 
Our model shows approximatively 70\% F1 score on both datasets with only 10 shots condition. This is a promising starting point since we could deploy CI tools based on such a model, and in subsequent iterations, instead of keeping manually labeling new data until there is enough data to start, developers would obtain more training data by only reviewing the predicted labels and correcting the wrong ones.
As expected, the performance of our model increases as the amount of training data increases, and it has already shown strong competitiveness in the case of 50shot. It reached an F1 score of 80.06\% on Dataset I and 76.41\% on Dataset II, respectively. Generally, baselines based on a larger volume of pre-training models can achieve better results. 
However, we also found that the introduction of code change features did not achieve an ideal improvement, which may be due to the code change information being collected from the commit log, usually, the change is not large, as in dataset I. And it is difficult to capture enough information from sentence-level commit representations. 
\begin{figure*}[ht]
  \centering
  \includegraphics[width=\linewidth]{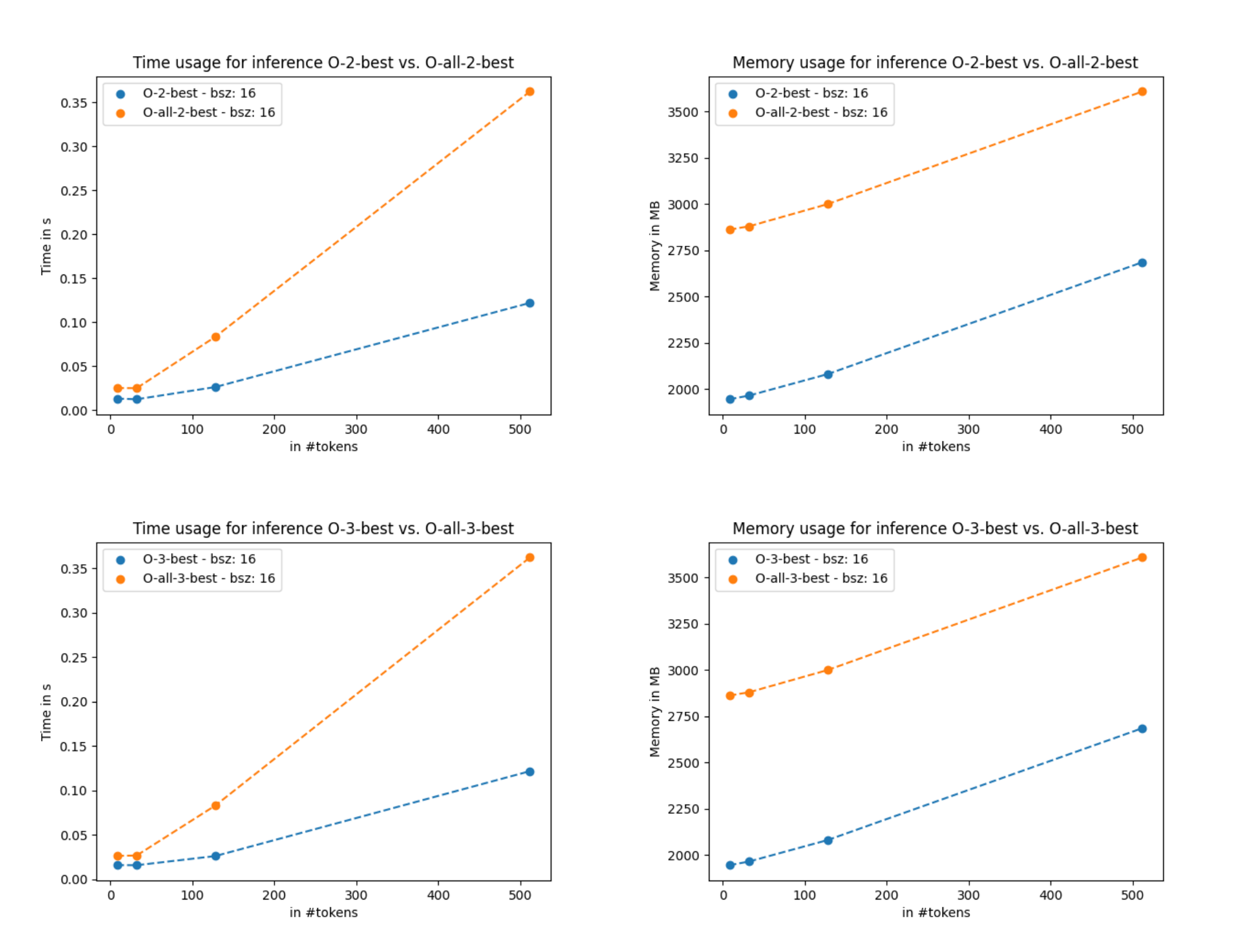}
  \caption{Computational Cost on Two Datasets
  }
  \label{fig: Computational Cost on Two Datasets}
\end{figure*}
This situation is more obvious on Dataset II. The code change features of Dataset II are vectorized information processed by experts, which is more effective as additional supplementary information when the amount of data is disclosed, and the performance on the {5, 10, 15} shots scene has verified this idea. However, when there is sufficient commit message information, it may lead to low discrimination or even affect the accuracy of the model because it is too sparse. We can observe that in the case of 50 shots in Dataset II, compared to $BooCC_{mge}$,  the accuracy of $BooCC_{mge+ cc}$ has dropped about 0.1\%.
Further, we observed the overall trend of ACC performance of the four basslines in different shots, and we visualized the results as Fig. \ref{fig: Fewshot perfermance}. We can see that in the case of fewshot scenario, The baselines based on paraphrase-mpnet-base-v2 have achieved better results on both datasets. In particular, on dataset I, the all-roberta-large-v1-based baselines $BooCC_{mge}$ and $BooCC_{mge+ cc}$ consistently outperform the other baselines, finally reaching an ACC of 80.01.
In dataset II, the effect of the model fluctuated slightly. Before 20shots, $BooCC_{mge}$ and $BooCC_{mge+ cc}$ achieved good effect, and after 20 shots, $BooCC_{mge}^{RoBERTa}$ based on all-roberta-large-v1 achieved the best score, 76.36\% Acc. As can be seen from the dotted lines, the methods with added code change features, the change of ACC is almost indistinguishable from other baselines, in Dataset I and Dataset II.

Overall, our model achieves satisfactory performance on the few-shot scenes of the two datasets, which not only demonstrates the significant impact of contrastive learning-based methods on improving submission classification tasks, but also shows that our model has strong adaptability to different data. 

\subsection{The Computational costs of BooCC}
\label{sec: Efficiency of Booc Computational costs}
In the previous section, we noticed that the baselines based on paraphrase-mpnet-base-v2 are better than all-roberta-large-v1-based baselines in some cases. In real development scenarios, the code maintenance cycle is often very short, and a large number of codes are submitted every day. Therefore, the inference time and size of the CC model are very important for deployment. We now discuss the inference time and model size of our several baselines to choose the most suitable one in this section.
\begin{table}[!ht]
  \centering
  \caption{Computational Cost on Dataset I}
  
  \begin{tabular}{lllll}
    \hline
      Model & Sequence\_Length & Inference Time & Memory \\ \hline
      O-2-best & 8 & 0.0131 & 1945 \\ 
      O-2-best & 32 & 0.0123 & 1965 \\ 
      O-2-best & 128 & 0.0262 & 2081 \\ 
      O-2-best & 512 & 0.1219 & 2685 \\ 
      O-all-2-best & 8 & 0.0251 & 2861 \\ 
      O-all-2-best & 32 & 0.0249 & 2879 \\ 
      O-all-2-best & 128 & 0.0833 & 2999 \\ 
      O-all-2-best & 512 & 0.3622 & 3607 \\ \hline
  \end{tabular}
  \label{tab: Computational Cost on Dataset I}
\end{table}

\begin{table}[!ht]
  \centering
  \caption{Computational Cost on Dataset II}
  \begin{tabular}{llllll}
  \hline
    Model & Sequence\_Length & Inference Time & Memory \\ \hline
    O-3-best & 8 & 0.016 & 1945 \\ 
    O-3-best & 32 & 0.0159 & 1965 \\ 
    O-3-best & 128 & 0.0262 & 2081 \\ 
    O-3-best & 512 & 0.1217 & 2685 \\ 
    O-all-3-best & 8 & 0.0264 & 2861 \\ 
    O-all-3-best & 32 & 0.0268 & 2879 \\ 
    O-all-3-best & 128 & 0.083 & 2999 \\ 
    O-all-3-best & 512 & 0.3621 & 3607 \\ \hline
  \end{tabular}
  \label{tab: Computational Cost on Dataset II}
\end{table}
We tested the performance of methods
based on paraphrase-mpnet-base-v2 and  all-roberta-large-v1-based pre-training models $BooCC_{mge}$ and $BooCC_{mge}^{RoBERTa}$ under different sequence lengths. 
We set sequence length to \{8, 32, 128, 512\}. From Tab.\ref{tab: Computational Cost on Dataset I} and Tab.\ref{tab: Computational Cost on Dataset II}, not surprisingly, we can see that the computational cost of $BooCC_{ mge}$ is significantly lower than $BooCC_{mge}^{RoBERTa}$ on both datasets, which is in line with expectations, because the more parameters the model uses, the longer the inference time of the model will be, and the size of the model will be larger.
In order to compare the benchmarks of the models more intuitively, we visualized the experimental results. It can be seen from Fig.\ref{fig: Computational Cost on Two Datasets} that the blue curve of $BooCC_{mge}$ is throughout lower than the orange curve of $BooCC_{mge}^{RoBERTa}$.
According to the previous experiments, we can conclude that only relying on paraphrase-mpnet-base-v2 pre-trained ST as the ENcoder, and the $BooCC_{mge}$ model using mge features is the most cost-effective model. This also verifies that our proposed method is simple but effective in solving CC problems.


\section{Conclusion} 

Commit Classification(CC) is an important task in software maintenance. 
Existing models need lots of manually labeled data for fine-tuning process, and ignore essential sentence-level semantic information for discovering the difference between diverse commits.
In this work, we propose a contrastive learning based framework for commit classification, which self-supervised generates $K$ sentences and pseudo-labels according to the labels of the dataset. This method randomly groups the augmented data $N$ times to compare the commit's similarity with the positive $T_p^{|C|}$ and negative $T_n^{|C|}$ samples, by introducing the sentence-level commits representations. 
Experimental results on two public datasets demonstrate that our proposed method has better adaptability, and can simply but effectively distinguish different commits with only a fewshot samples for training. We public the dataset, code, and experimental results  in \url{https://github.com/AppleMax1992/CommitFit}.
In future work, we are going to investigate knowledge distillation methods to further reduce the model size. Or explore utilizing the prompt-based method to fuse external knowledge to solve the CC problem.
\section*{CRediT authorship contribution statement}
\textbf{Jiajun Tong}: Methodology, Software, Writing, original draft, Editing.
\textbf{Zhixiao Wang}: Supervision, Writing - review \& editing. 
\textbf{Xiaobin Rui}:
Writing - review \& editing.

\section*{Declaration of competing interest}
The authors declare that there is no conflict of interest regarding the publication of this article.

\section*{Data availability}
The raw/processed data required to reproduce these findings cannot be shared at this time as the data also forms part of an ongoing study.

\section*{Acknowledgments}
This work was supported by the National Natural Science Foundation of
China (No. 61876186) and the Xuzhou Science and Technology Project (No.
KC21300).

\bibliographystyle{cas-refs}
\bibliography{cas-refs}

\end{document}